\documentclass[aps,pra,10pt,twocolumn,showpacs,nolongbibliography,superscriptaddress]{revtex4-2}
\usepackage{physics}
\usepackage{graphicx}
\usepackage[colorlinks=true,linkcolor=blue,citecolor=red,plainpages=false,pdfpagelabels]{hyperref}
\usepackage{times}
\usepackage{amssymb}
\usepackage{amsthm}

\usepackage{mathtools}

\begin{document}
\title{Effect of measurements on quantum speed limit}

\author{Abhay Srivastav}
\email{abhaysrivastav@hri.res.in}
\author{Vivek Pandey}
\affiliation{ Quantum Information and Computation Group,\\
Harish-Chandra Research Institute, A CI of Homi Bhabha National Institute, Chhatnag Road, Jhunsi, Prayagraj 211019, India
}
\author{Arun K Pati}
\affiliation{ Centre for Quantum Science and Technology (CQST),\\
International Institute of Information Technology, Hyderabad-500032, India}

\begin{abstract}
Given the initial and final states of a quantum system, the speed of transportation of state vector in the projective Hilbert space governs the quantum speed limit. Here, we ask the question what happens to the quantum speed limit under continuous measurement process. We model the continuous measurement process by a non-Hermitian Hamiltonian which keeps the evolution of the system Schr{\"o}dinger-like even under the process of measurement. Using this specific measurement model, we prove that under continuous measurement, the speed of transportation of a quantum system tends to zero. Interestingly, we also find that for small time scale, there is an enhancement of quantum speed even if the measurement strength is finite. Our findings can have applications in quantum computing and quantum control where dynamics is governed by both unitary and measurement processes.
\end{abstract}

\maketitle
{\bf{Introduction}}--
The laws of quantum mechanics impose a fundamental bound on the time scale for the evolution of a quantum system. Mandelstam and Tamm were the first to realise this in the context of unitary evolution \cite{Mandelstam1945}. They proved that this time scale is the minimum time required for the system to evolve from an initial state to some final state and gave a clear interpretation of the time-energy uncertainty relation. This was the first instance of quantum speed limit (QSL), a notion, that sets a lower bound on the time required to evolve between two quantum states under a given dynamics. Anandan and Aharonov then geometrised quantum evolutions and introduced the concept of the evolution speed of a system using the Fubini-Study metric, formalising the notion of QSL in the projective Hilbert space \cite{Anandan1990}. Using the Riemannian metric, the distance function was defined and the speed of transportation was also introduced in Ref.\cite{Pati_1991}. An alternative bound to the rate of evolution was derived by Margolus and Levitin based on the expectation value of the Hamiltonian of the system \cite{Margolus1998}. The QSLs have been extensively studied for the unitary evolution of pure as well as mixed states \cite{Bhattacharyya_1983,Vaidman1992,Uhlmann1992,Pfeifer1993,Levitin2009,Deffner_2013,Mondal2016,Mondal_2016,Campaioli2018,Shao2020,Ness2021,Ness2022,Bagchi2022}. They have also been generalised for non-unitary evolution \cite{Bender2007,Uzdin_2012,delCampo2013,Taddei2013,Deffner2013,Pires2016,Campaioli2019,Brody2019,Das2021,Impens2021,Lan_2022,Nakajima_2022,Thakuria_2024} as well as many-body systems \cite{Giovannetti_2004,Batle2005,Zander_2007,Borras_2008,Brouzos2015,Bukov2019}. Recently, a topological speed limit was derived in Ref.\cite{Van2023} providing useful insights into the evolution speed from topological perspective. An exact speed limit for unitary evolution of $d$-dimensional quantum systems was derived in Ref.\cite{Pati2023}. In recent years, QSLs have also been studied for observables \cite{Pintos2022,Hamazaki2022,Mohan2022,Hornedal2023} and for correlation and informational measures \cite{Mohan_2022,Pandey2023}. Using the stronger uncertainty relations \cite{Maccone2014}, the stronger quantum speed limit for state has been proved in Ref.\cite{thakuria2022} and the stronger quantum speed limit for observable has been proved in Ref. \cite{Divyansh2024}.

Even though unitary quantum dynamics is well understood, the quantum measurement problem still continues to be a subject of topical interest. Within the von Neumann model, one can describe the measurement process as a coupling between the system and the apparatus where the combined system evolves according to the Schr\"odinger equation. After the interaction, the apparatus reads one of the eigenvalues of the system observable with full
accuracy. The state of the system which is typically in a superposition of different eigenstates corresponding to different
eigenvalues of the observable, collapses to that eigenstate corresponding to the eigenvalue being measured. This
type of measurement is called the precise and instantaneous one. The notion of repeated measurements on a quantum system can give rise to counterintuitive effects such as the quantum Zeno effect and quantum Zeno phase effect \cite{Misra1977,Itano1990,Braginsky1995,Pati_1996,Beige_1997,Erik1997,Pati_1998,Pati1999,Fischer2001,Jonas2003,Faria_2003,Schmidt2004,Itano_2009}. There have been several approaches to describe the continuous measurement using non-unitary dynamics \cite{Jacobs2014}. For example, a proposal of Kulaga, where the measurement of some observable of a system is described by a non-unitary evolution equation \cite{Kulaga1995}. The evolution equation is Schr\"odinger-like except
that there is an imaginary term in the Hamiltonian that takes into account the process of measurement. The
evolution is thus non-unitary which signifies the irreversibility of the measurement process. The Kulaga model of measurement captures the essential features of the continuous measurement process. Using this model, one can prove the usual quantum Zeno effect as well as a modified quantum Zeno paradox. In addition, one can prove the quantum Zeno phase effect which says that under continuous measurement the system does not acquire any geometric phase \cite{Pati_1996,Pati1999}.

In this paper, we address the following question: what happens to the quantum speed limit under continuous measurement process. To motivate our question, imagine that there is a runner on the playground with initial and final positions fixed. The time the player will take to complete the run is essentially governed by how fast the player can run on the ground. But, suppose that there is another person who is trying to pull the player from behind. Then, the speed of the player will be governed by how frequent is the disturbance and with what strength the player is being pulled from behind. Now, we translate this scenario to the quantum world where we have a free evolution of the quantum system by a Hermitian Hamiltonian and then the system is being continuously disturbed by a measurement process which is governed by the non-Hermitian part of the Hamiltonian. Without the measurement part, 
given the initial and the final states of a quantum system, the speed of transportation of state vector in the projective Hilbert space governs the quantum speed limit. Then, the natural question that comes to mind is that what happens to the quantum speed under measurement? Using a specific measurement model where the process of measurement is modeled by a non-Hermitian Hamiltonian, we prove that under continuous measurement, the speed of transportation of a quantum system  will tend to zero. Unlike the classical case, here, we find that for small time scale, there is an enhancement of quantum speed even if the measurement strength is finite. Our findings can have applications in quantum computing and quantum control where dynamics is governed by both unitary and measurement processes. Note that with a different motivation, Ref.\cite{Pintos_2019} studied quantum speed limits under the influence of continuous measurements. They considered a measurement model that rendered the evolution of the system stochastic and found that the evolution speed depends on the record of the measurement outcomes. It was also inferred that there are trajectories for which standard QSL can be violated and continuous quantum measurements
can induce Brownian dynamics in the Hilbert space. However, it has not been proved yet, that under continuous measurement the speed of transportation of a quantum system can vanish. In the present work, the measurement model that we consider, however, keeps the evolution Schr\"odinger-like and is not stochastic type. This model is
better suited to study the dynamical effects of measurements on the geometry of the state space and prove the main result of this letter. Since the measurement model that we consider in this article to arrive at our results models continuous measurements by non-Hermitian Hamiltonian, in the following, we first discuss the dynamics of a quantum system that is generated by a general non-Hermitian Hamiltonian.\\

{\bf{Preliminaries}}--
Consider a $d$-dimensional quantum system initially prepared in the state $\ket{\Phi(0)} \in \mathcal {H}$. The dynamics of the system is governed by a non-Hermitian Hamiltonian $H$ which can be decomposed as
\begin{equation}
    H=H_0 - i H_1,
\end{equation}
where $H_0$ and $H_1$ are Hermitian operators. The state $\ket{\Phi(t)}$ evolves according to the modified Schr{\"o}dinger's equation as given by
\begin{equation}\label{seq}
    i \hbar \frac{d}{dt}\ket{\Phi(t)}=(H_0 - i H_1)\ket{\Phi(t)}.
\end{equation}
Since the Hamiltonian is non-Hermitian the corresponding time evolution operator is non-unitary. Therefore, the norm of the state is not preserved in time. We can define a state $\ket{\Psi(t)}$ that is normalized at all times as
\begin{equation}\label{psidef}
    \ket{\Psi(t)}:=\frac{\ket{\Phi(t)}}{\norm{\ket{\Phi(t)}}},
\end{equation}
where $\norm{\ket{\Phi(t)}}=\sqrt{\innerproduct{\Phi(t)}}$. Using Eq.\eqref{seq} we see that the dynamical equation for $\ket{\Psi(t)}$ is
\begin{equation}\label{nseq}
    i \hbar \frac{d}{dt}\ket{\Psi(t)}=H_0\ket{\Psi(t)}-i (H_1-\bra{\Psi(t)}H_1\ket{\Psi(t)})\ket{\Psi(t)}.
\end{equation}
The state space of a quantum system with Hilbert space ${\cal H}$ is the projective Hilbert space $\mathcal{P}(\cal{H})$ which is defined as $\mathcal{P}{(\cal H)}:=({\cal H}-\{0\})/\sim$ where $\ket{\phi}\sim\ket{\psi}$ iff $\ket{\phi}=c\ket{\psi}$, $c\in\mathbb{C}-\{0\}$. Each point of the projective Hilbert space corresponds to a distinct physical state. Now we know that $\mathcal{P}{(\cal H)}$ is equipped with a natural metric called the Fubini-Study metric. For arbitrary time-continuous dynamics, the generalised Fubini-Study metric is given by \cite{Pati1995,Pati_1991}
\begin{equation}
    dS^2=4[\langle\dot{\Psi}(t)|\dot{\Psi}(t\rangle)-(i\langle\Psi(t)|\dot{\Psi}(t)\rangle)^2]dt^2,
\end{equation}
where $\ket{\Psi(t)}=\ket{\Phi(t)}/\norm{\ket{\Phi(t)}}$ as defined above. It is a generalisation of the metric given in Ref.\cite{Anandan1990}, where the metric was invariant under a transformation of the type $\ket{\Phi(t)}\rightarrow e^{i\alpha(t)}\ket{\Phi(t)}$. The above metric, however, is invariant under generalised local gauge transformations of the type $\ket{\Phi(t)}\rightarrow \chi(t)e^{i\alpha(t)}\ket{\Phi(t)}$.
Using the above equation we can write the evolution speed of a system in $\mathcal{P}(\cal{H})$ as 
\begin{equation}
    V=\frac{dS}{dt}=2\sqrt{\langle\dot{\Psi}(t)|\dot{\Psi}(t\rangle)-(i\langle\Psi(t)|\dot{\Psi}(t)\rangle)^2}.
\end{equation}
Now the equation of motion for $\ket{\Psi(t)}$ implies
\begin{eqnarray}
 \langle\dot{\Psi}(t)|\dot{\Psi}(t\rangle)&=&\frac{1}{\hbar^2}[\bra{\Psi(t)}H_0^2\ket{\Psi(t)}+\bra{\Psi(t)}\Delta H_1^2\ket{\Psi(t)}\nonumber\\
 &&+i \bra{\Psi(t)}[H_1,H_0]\ket{\Psi(t)}],   
\end{eqnarray}
and the overlap of $\ket{\Psi(t)}$ and $|\dot{\Psi}(t)\rangle$ is
\begin{eqnarray}
 \langle\Psi(t)|\dot{\Psi}(t\rangle)=-\frac{i}{\hbar}[\bra{\Psi(t)}H_0\ket{\Psi(t)}]. 
\end{eqnarray}
Using the above three equations we get the evolution speed as \cite{Brody2012,Impens2021,Thakuria_2024}
\begin{equation}\label{speed}
    V=\frac{2}{\hbar}\sqrt{\Delta H_0^2 + \Delta H_1^2 + i \langle[H_1,H_0]\rangle}.
\end{equation}
Note that due to the non-Hermiticity of the Hamiltonian, the evolution speed depends on time. Therefore, the important quantity to consider is the time-averaged evolution speed defined as
\begin{equation}\label{time-averaged speed}
    \overline{V}=\frac{1}{T}\int_0^T V dt,
\end{equation}
where T is the total time of evolution. In the projective Hilbert space $\mathcal{P}(\cal{H})$ using the Fubini-Study metric we can measure the distance between any two states called as the Fubini-Study distance $S$ in terms of which the time of evolution can be written as
\begin{equation}
    T=\frac{S}{\overline{V}},
\end{equation}
where $\overline{V}$ is the time-averaged evolution speed as defined above. Since the shortest path between any two states is the geodesic, the time taken for the evolution will be minimum if the state evolves along the geodesic. Hence, the time of the evolution is lower bounded as \cite{Thakuria_2024}
 \begin{equation}\label{QSL}
     T \geq T_{\rm QSL}= \frac{S_{0}}{\overline{V}},
 \end{equation}
where $T_{QSL}$ is called the quantum speed limit time and $S_{0}=2\cos^{-1}({|\langle\Psi(0)|\Psi(T)\rangle|})$ is the geodesic distance between the states $\ket{\Phi(0)}$ and $\ket{\Phi(T)}$. The bound given above applies to arbitrary time-continuous dynamics and therefore can be used to study the effect of time-continuous measurements on the evolution speed.\\  

{\bf{Effect of measurements on quantum speed}}--
Consider a $d$-dimensional quantum system in the state $\ket{\Phi(t)}$ and an observable $A$ whose measurements are to be performed on the system. The measurement setup induced by the observable $A$ is described by a POVM (Positive Operator Valued Measure) which is a collection of positive operators $\{\Pi_i\}$ with $\sum_i\Pi_i=\mathbb{I}$, where the label $i$ denotes the $i$-th outcome and the probability of obtaining that outcome is $p_i=\braket{\Phi(t)}{\Pi_i|\Phi(t)}$. Let a series of measurements of the observable $A$ is performed on the system in the state $\ket{\Phi(t)}$ and $\delta t$ is the time duration between those measurements. Then the limit $\delta t\rightarrow0$ describes the process of continuous measurements. However, it is not easy to take this limit in practice. Note that during the process of measurement, the system first interacts with a quantum probe which is initially prepared in some state and then the measuring device measures some observable of the probe. Therefore, to completely describe any specific measurement the knowledge about the probe is required. However, we can still discuss the most general features of the measurement process by the knowledge about the observable that is to be measured and some parameters of the measuring device. The continuous measurement model that we discuss here does exactly that where the concerned parameters are associated with the frequency and the accuracy of the measurements. The model describes the continuous measurement of some observable of the system by a Schr{\"o}dinger-like evolution equation with the corresponding total Hamiltonian being non-Hermitian \cite{Kulaga1995}. The total Hamiltonian of the system is given as $H=H_0-iH_1$, where $H_0$ is the free Hamiltonian and $H_1$ describes the measurement process. The non-Hermitian Hamiltonian induces a non-unitary time evolution operator which takes into account the irreversible evolution of the system due to measurements. The non-Hermitian part of the Hamiltonian which describes the measurement process is given by
\begin{equation}\label{measureham}
    H_1=\hbar f g\left(\frac{A-\tilde{a}(t)\mathbb{I}}{\Delta \tilde{a}}\right),
\end{equation}
where $A$ is the observable being measured, $\tilde{a}(t)$ is the measurement outcome given by the measuring apparatus continuously in time. The measuring apparatus is characterized by two parameters $f$ and $\Delta \tilde{a}$, where $f$ and $\Delta \tilde{a}$ determine the frequency and the accuracy of the measurements, respectively. The function $g$ satisfies: $g(x)\geq0$ and $g(0)=0$. It is important to understand the physical significance of the states $\ket{\Phi(t)}$ and $\ket{\Psi(t)}$ in this measurement model. Let $\tilde{a}(\tau)$, $0\leq\tau\leq t$ be the measurement results obtained for the time duration $t$. Let $\{a_i\}$ be the non-degenerate finite spectrum of the observable $A$ with the corresponding eigenvectors being $\{\ket{a_i}\}$. Note that the measurement results do not necessarily belong to the spectrum of the observable due to the inaccuracy of the measuring apparatus characterized by $\Delta \tilde{a}$. The vector $\ket{\Phi(t)}$ can be expressed in the basis $\{\ket{a_i}\}$ as
\begin{equation}
 \ket{\Phi(t)}=\sum_i \Phi_i(t)\ket{a_i}.   
\end{equation} 
Using the definition of $\ket{\Psi(t)}$ (see Eq. \eqref{psidef}) and the above equation we have $\abs{\Phi_i(t)}^2=\abs{\Psi_i(t)}^2\norm{\Phi(t)}^2$ where $\Psi_i(t)$ is the $i$-th expansion coefficient of $\ket{\Psi(t)}$ in the basis $\{\ket{a_i}\}$. The probability of getting the outcome $a_i$ if a precise instantaneous measurement is done at time $t$ is $\abs{\Psi_i(t)}^2$  and $\norm{\Phi(t)}^2$ is the probability  of obtaining the results $\tilde{a}(\tau)$, $0\leq\tau\leq t$. So, $\ket{\Phi(t)}$ describes both the state of the system and the probability of obtaining particular results of the measurements and $\ket{\Psi(t)}$ describes just the state of the system \cite{Kulaga1995}. We will now use the formalism discussed above to study the effect of measurements on the evolution speed of the system. The time evolution operator corresponding to the Hamiltonian $H=H_0-iH_1$ is given by
\begin{eqnarray}\label{unitary}
    U(t)=\exp\left[-\frac{i}{\hbar}\int(H_0-iH_1)dt\right].
\end{eqnarray}
We now assume that the measurement process is frequent enough, i.e., $f$ is sufficiently large so that the free evolution of the system can be neglected then $U(t)\approx\exp[-\int dt H_1/\hbar ]$.
Then, the dynamical equation for $\ket{\Phi(t)}$ gives
\begin{eqnarray}
    \ket{\Phi(t)}&=&U(t)\ket{\Phi(0)}\nonumber\\
    &=&\exp[-f\int g\left(\frac{A-\tilde{a}(t)\mathbb{I}}{\Delta \tilde{a}}\right)dt]\ket{\Phi(0)},\nonumber
\end{eqnarray}
where $\ket{\Phi(0)}$ is the initial system of the system. Using the above equation we can write the equation of motion for the amplitude $\Phi_i(t)$ as
\begin{equation}
\Phi_i(t)=\exp[-f\int g\left(\frac{a_i-\tilde{a}(t)}{\Delta \tilde{a}}\right)dt]\Phi_i(0).\nonumber
\end{equation}
Since the strength of the measurement is sufficiently large the evolution speed of the quantum system is predominantly given by the variance of $H_1$. Now the expectation value of $H_1$ in the normalized state $\ket{\Psi(t)}$ is
\begin{equation}
\bra{\Psi(t)}H_1\ket{\Psi(t)}=\frac{\bra{\Phi(t)}H_1\ket{\Phi(t)}}{\innerproduct{\Phi(t)}}.
\end{equation}
Using the form of the time-evolved amplitude $\Phi_k(t)$ the above equation can be written as
\begin{widetext}
\begin{eqnarray}\label{expH}
\bra{\Psi(t)}H_1\ket{\Psi(t)} =\frac{\sum_i\exp[-2f\int g(\frac{a_i-\tilde{a}(t)}{\Delta \tilde{a}})dt]\hbar fg(\frac{a_i-\tilde{a}(t)}{\Delta \tilde{a}}) \abs{\Phi_i(0)}^2}{\sum_i\exp[-2f\int g(\frac{a_i-\tilde{a}(t)}{\Delta \tilde{a}})dt]\abs{\Phi_i(0)}^2}.
\end{eqnarray}
\end{widetext}
Let us consider the limiting case where the strength of the measurements $f$ tends to infinity. In this limit, the exponentials in the finite sums decay rapidly. Let $r$ be the number such that $x_i=\int g\left(\frac{a_i-\tilde{a}(t)}{\Delta \tilde{a}}\right)dt$ is minimum for $i=r$, i.e., $x_r=\min\{x_i\}$. Then the above equation reduces to
\begin{eqnarray}
\lim_{f\rightarrow\infty} \bra{\Psi(t)}H_1\ket{\Psi(t)} &=&\frac{\exp[-2fx_r]\hbar fg(\frac{a_r-\tilde{a}(t)}{\Delta \tilde{a}})\abs{\Phi_i(0)}^2}{\exp[-2fx_r]\abs{\Phi_i(0)}^2}\nonumber\\
&=&\hbar fg\left(\frac{a_r-\tilde{a}(t)}{\Delta \tilde{a}}\right).
\end{eqnarray}
Similarly, we get
\begin{eqnarray}
\lim_{f\rightarrow\infty} \bra{\Psi(t)}H_1^2\ket{\Psi(t)} =\left(\hbar fg\left(\frac{a_r-\tilde{a}(t)}{\Delta \tilde{a}}\right)\right)^2.
\end{eqnarray}
Hence, the variance of $H_1$ tends to zero as the strength of the measurements tends to infinity. Since we have assumed that in this limit the free evolution of the system can be neglected therefore we see that evolution speed of the system vanishes in the limit $f\rightarrow\infty$. Our analysis shows that if the strength of the measurements is sufficiently large such that the effect of the measurements overcomes the free evolution of the system then the initial state gets evolved to the state $\ket{a_r}$ in very small time, where the index $r$ is determined by the parameters of $H_1$. Once the system reaches this state, its further evolution is then frozen. We will now show that in the small time limit if in addition to being evolved by the free Hamiltonian $H_0$, a quantum system is also being measured continuously for an observable $A$ that commutes with the free Hamiltonian then there is an increment in the evolution speed of the system under suitable conditions.\\

{\bf{Measurement induced speed-up in evolution}}--
If the observable $A$ commutes with the free Hamiltonian $H_0$, then the time evolution operator becomes $U(t)=\exp[-\frac{iH_0 t}{\hbar}]\exp[-f\int g\left(\frac{A-\tilde{a}(t)\mathbb{I}}{\Delta \tilde{a}}\right)dt]$. Therefore, if the initial state of the system is $\ket{\Phi(0)}=\sum_i\Phi_i(0)\ket{a_i}$ the time-evolved amplitude $\Phi_i(t)$ is given by
\begin{equation}
\Phi_i(t)=\exp[-\frac{iE_{0i}t}{\hbar}]\exp[-f\int g\left(\frac{a_i-\tilde{a}(t)}{\Delta \tilde{a}}\right)dt]\Phi_i(0),\nonumber
\end{equation}
where $\{a_i\}$ and $\{E_{0i}\}$ are the eigenvalues of $A$ and $H_0$, respectively.
From now on  we work under the assumption that the measurement results are constant in time $\tilde{a}(t)=\tilde{a}$ for all $t$. Let us now consider the limiting case where the factor $2f\int g(\frac{a_i-\tilde{a}(t)}{\Delta \tilde{a}})dt=2fg(\frac{a_i-\tilde{a}}{\Delta \tilde{a}})t$ concerning the parameters describing the measurement and the time of the evolution is small enough so that the approximation $\exp[-2fg(\frac{a_i-\tilde{a}}{\Delta \tilde{a}})t]\approx 1-2fg(\frac{a_i-\tilde{a}}{\Delta \tilde{a}})t$ is meaningful. By making this approximation, we would like to see what happens if the system is subjected to continuous measurements for a small time $t$. The variance of $H_0$ in the state $\ket{\Psi(t)}$ under the said approximation is (see Appendix \ref{detailedproof})
\begin{eqnarray}
   \Delta H_0^2(t)&=& \Delta H_0^2(0)+2tf[ \langle H_0^2 \rangle_0 \langle g \rangle_0- \langle g H_0^2 \rangle_0\nonumber\\
   &&-2{\langle H_0 \rangle_0^2} \langle g \rangle_0+2\langle g H_0 \rangle_0 \langle H_0 \rangle_0],
\end{eqnarray}
where $\Delta H_0^2(0)$ is the variance of $H_0$ in the initial state $\ket{\Psi(0)}$, $\langle O\rangle_0$ denotes the expectation value of the operator $O$ in $\ket{\Psi(0)}$ and we have neglected terms of the order $\mathcal{O}(f^2g^2)$. The variance of $H_1$ vanishes under this approximation (see Appendix \ref{detailedproof}). Therefore, using Eq.\eqref{speed} and the fact that $[H_1,H_0]=0$ the evolution speed of the system is given by
\begin{eqnarray}
    V=\frac{2}{\hbar}\sqrt{\Delta H_0^2(0)+2tfX},
\end{eqnarray}
$X=2\langle H_0 \rangle_0 \text{Cov}(g,H_0)-\text{Cov}(g,H_0^2)$ and $\text{Cov}(A,B)=\langle AB \rangle-\langle A \rangle\langle B \rangle$ is the covariance of $A$ and $B$. In the case of no measurements, the evolution speed is given by the variance of the free Hamiltonian in the initial state, $\Delta H_0^2(0)$. Thus, we see from the above equation that for small times of evolution, continuous measurement provides an unexpected increment in the evolution speed of the system compared to the speed under free evolution if $X>0$. Note that a similar speed-up in the evolution of the state has also been observed for non-Markovian dynamics \cite{Cimmarusti2015,Deffner2013}. In our case, the continuous measurements make the dynamical equation non-linear (see Eq.\eqref{nseq}) and thus the observed speed-up must be attributed to this non-linearity. In this way, our work can be considered as complementary to that of non-Markovian speed-ups. For a different approach to non-linear speed-up of the evolution see Ref.\cite{Deffner_2022}. Now, since the speed depends on time due to the non-Hermiticity of the total Hamiltonian, we can calculate the time-averaged speed as
 \begin{equation}
    { \overline{V}=\frac{2}{\hbar T} \int^T_0 {\sqrt{\Delta H_0^2(0)+2tfX}}dt},
\end{equation}
where $T$ is the total time of evolution. In the following, we will elaborate our results with an example.

\begin{figure}[htp]
    \centering
    \includegraphics[width=6.6cm]{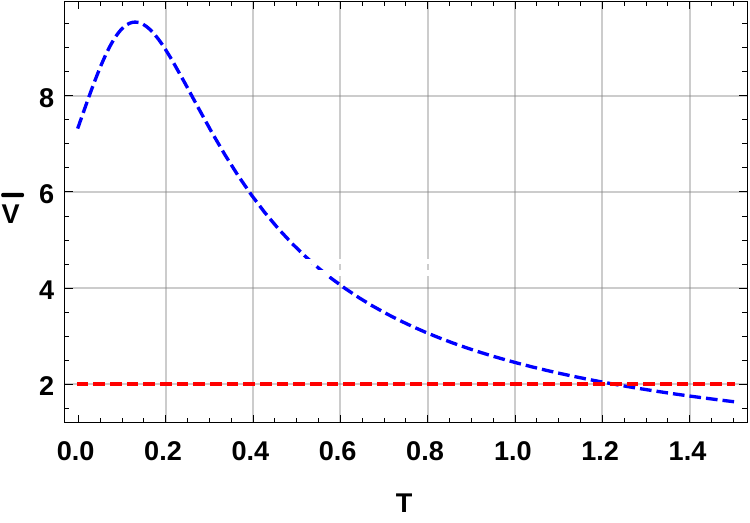}
  \caption{Time-averaged evolution speed as a function of the total time of evolution. The eigenvalues of the observable $A$ are $(a_1,a_2)=(0.03,0.05)$, the accuracy of the measuring apparatus is $\Delta \tilde{a}=0.01$ and the measurement results are constant in time $\tilde{a}(t)=0.02$. The blue line corresponds to $f=5$ and the red line corresponds to $f=0$.}
    \label{avgspeedvstime}
\end{figure}

\begin{figure}[htp]
    \centering
    \includegraphics[width=6.6cm]{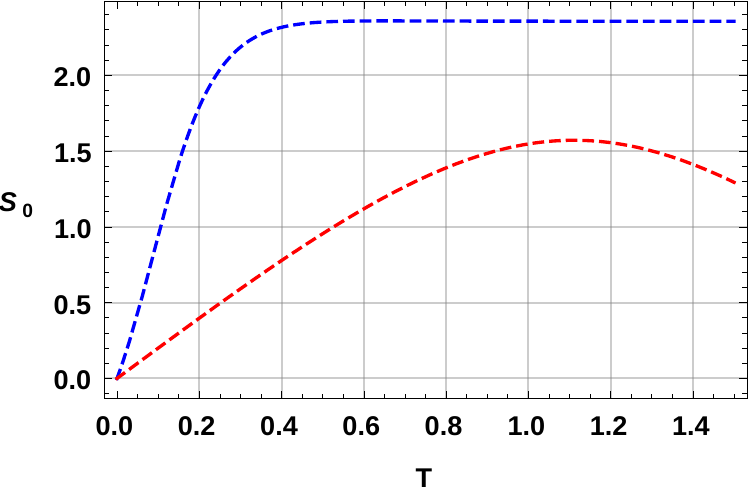}
  \caption{Distinguishability of the initial and the time-evolved state as a function of the total time of evolution. The blue line corresponds to $f=5$ and the red line corresponds to $f=0$.}
    \label{Geodistancevstime}
\end{figure}

Let us consider a spin half particle with Hamiltonian $\omega\sigma_z$ in an external magnetic field along the $x$-axis. Then the total free Hamiltonian is
\begin{equation}\label{freeH}
    H_0=H_{particle}+H_{field}=\omega\sigma_z+\alpha\sigma_x,
\end{equation}
where we put $\omega=\alpha=\hbar$ for simplicity. Let the observable $A$ that is measured continuously and is diagonal in the eigenbasis of $H_0$ be
\begin{equation}\label{measuredobs}
    A=\frac{a_{21}}{2\sqrt{2}}\left(
\begin{array}{cc}
 \sqrt{2}(\frac{\tr(A)}{a_{21}})+1 & 1 \\
 1 & \sqrt{2}(\frac{\tr(A)}{a_{21}})-1 \\
\end{array}
\right),
\end{equation}
where $a_{21}=a_2-a_1$ and $a_1$, $a_2$ are the eigenvalues of $A$. Now choosing $g(x)=\frac{1}{4}x^2$ as in Ref.\cite{Kulaga1995}, we can rewrite $H_1$ using Eq.\eqref{measureham} as
\begin{equation}
    H_1=\frac{\hbar f}{4\Delta\tilde{a}^2} (A^2+\tilde{a}(t)^2\mathbb{I}-2\tilde{a}(t)A).
\end{equation}
The non-unitary time evolution operator is then given as $U(t)=\exp[-i\frac{H_0t}{\hbar}]\exp[-\int \frac{H_1}{\hbar}dt]$ where we have used the fact that $[H_1,H_0]=0$. Let us now assume that $\tilde{a}(t)$ is a time independent quantity, i.e., $\tilde{a}(t)=\tilde{a}~\forall t$. If the initial state of the system is the maximally coherent state $\ket{\Psi(0)}=\frac{1}{\sqrt{2}}(\ket{0}+\ket{1})$, the time-evolved normalized state is given by
\begin{eqnarray}\label{finalstate}
    \ket{\Psi(t)}=U(t)\ket{\Psi(0)}=\frac{1}{N(t)}\left(z_1(t)\ket{0}+z_2(t)\ket{1}\right),\nonumber
\end{eqnarray}
where $N(t)=\sqrt{\abs{z_1(t)}^2+\abs{z_2(t)}^2}$ is the time-dependent normalization factor with the coefficients $z_1(t)=\frac{e^{-\delta t} \left[e^{\gamma t} \left(8 \left(2+\sqrt{2}\right) \Delta \tilde{a}^2-i \left(1+\sqrt{2}\right) p\right)+e^{\omega t} \left(8 \left(\sqrt{2}-2\right) \Delta \tilde{a}^2+\left(\sqrt{2}-1\right) i p\right)\right]}{2 \sqrt{2} \left(8 \sqrt{2} \Delta \tilde{a}^2-i p\right)}$ and $z_2(t)=\frac{e^{-\delta t} \left(e^{\gamma t}+e^{\omega t}\right)}{2 \sqrt{2}}$,
where $p=f(2\tilde{a}-a_1-a_2)(a_1-a_2)$, $ \delta=\frac{f\left(2 \tilde{a}^2-2 \tilde{a} (a_1+a_2)+a_1^2+a_2^2\right)}{4 \Delta \tilde{a}^2}$, $\gamma=\frac{\left(f (\tilde{a}-a_1)^2-4 i \sqrt{2} \Delta \tilde{a}^2\right)}{4 \Delta \tilde{a}^2}$ and $\omega=\frac{\left(f (\tilde{a}-a_2)^2+4 \sqrt{2} i \Delta \tilde{a}^2\right)}{4 \Delta \tilde{a}^2}$.
\begin{figure}[htp]
    \centering
    \includegraphics[width=6.6cm]{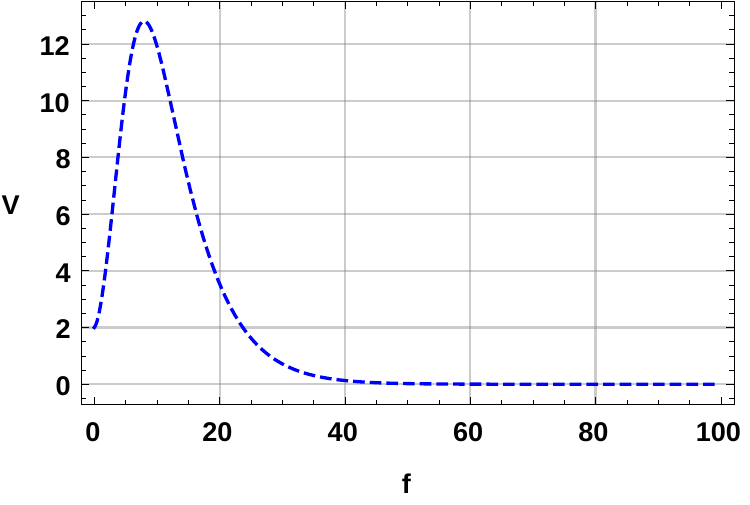}
  \caption{Evolution speed as a function of the strength of the measurements. The eigenvalues of the observable $A$ are $(a_1,a_2)=(0.03,0.05)$, the accuracy of the measuring apparatus is $\Delta \tilde{a}=0.01$, the measurement results are constant in time $\tilde{a}(t)=0.02$ and the speed is calculated at $t=0.1$.}
    \label{speedvsfrequency}
\end{figure}
Using the time-evolved state $\ket{\Psi(t)}$ we can now calculate the variances of the free Hamiltonian $H_0$ and the part governing the measurement process $H_1$ using which we plot the evolution speed numerically as a function of the strength of the measurements in Fig.\ref{speedvsfrequency}. We can see that at large $f$ the evolution speed decays very rapidly eventually vanishing completely. This elaborates our result that if the measurement process is fast enough it overcomes the free evolution and the evolution speed of the system vanishes. Note that for small $f$ there is a peculiarity in the plot showing an increase in the evolution speed by the virtue of performing measurements on the system. This shows an unexpected advantage of measurements in the evolution of quantum systems. To understand this advantage further let us fix the value of $f$ and plot the time-averaged evolution speed \eqref{time-averaged speed} as a function of the time of evolution Fig.\ref{avgspeedvstime}.
We can see that there is a clear measurement induced speedup in the evolution of the system compared to the case where there is no measurement. Note that this advantage in the speed will only be for a small time after which the effect of the measurements will stop the evolution completely where as in the absence of any measurement the evolution speed remains constant. This time depends on the parameters describing the measurement process. The increase in the evolution speed implies that the distinguishability of the initial state $\ket{\Phi(0)}$ and the time-evolved state $\ket{\Phi(T)}$ increases much faster if in addition to the free evolution, there is a continuous measurement being performed on the system. This is elaborated in Fig.\ref{Geodistancevstime} where the geodesic distance between the states $\ket{\Phi(0)}$ and $\ket{\Phi(T)}$, $S_{0}=2\cos^{-1}({|\langle\Psi(0)|\Psi(T)\rangle|})$ is taken as the measure of their distinguishability.
For $f=0$, $S_0$ increases slowly with time as the system evolves farther away from the initial state. It then reaches a peak where the system has been driven the farthest from the initial state $\ket{\Phi(0)}$ as much is possible by the free Hamiltonian $H_0$, after which the evolution of the system is such that its distance from the initial state starts decreasing. However, note that $S_0$ increases very sharply for $f=5$ compared to $f=0$ and then very quickly saturates to a fixed value after a small time. The sharp increase is due to the measurement induced increment in the evolution speed. And the quick saturation is due to the fact that the effect of the measurements overcomes that of the free evolution, by evolving the initial state to one of the eigenstates of the observable being measured, in a time that depends on the value of $f$. Since, we have assumed that $[H_1,H_0]=0$ there can be no further evolution once the system reaches this state. Moreover, note that for $f=0$ the maximum of $S_0$ is around 1.6 which gets surpassed in the case of $f=5$ in a small time. Therefore, if the system is being continuously measured in addition to being evolved by the free Hamiltonian it has access to states which are much farther than its initial state and to which it could never evolve by the free Hamiltonian alone.\\

{\bf{Conclusion}}--
In this work, we have studied the effect of measurements on the evolution speed of a quantum system. To do this we considered a measurement model where the process of measurement is described by a Schr{\"o}dinger-like evolution equation where the total Hamiltonian is non-Hermitian. We observed that if the measurement process is frequent enough so that the free evolution of the system can be neglected then in the limit $f\rightarrow\infty$ where $f$ determines the strength of the measurements, the evolution speed of the system vanishes. This is in sharp contrast to the standard quantum Zeno effect, where the transition to other state is prohibited. However, the speed of quantum system over a small time scale is never zero.  We also observed quite unexpectedly that under suitable conditions the evolution speed of the system can even increase as the parameter $f$ increases. Finally, we elaborated our results with an example and showed the measurement induced advantage in the dynamics of a quantum system. We believe that our results can have important applications in quantum computing and quantum control where dynamics is governed by both unitary and measurement processes.

{\it Acknowledgements:}
AS and VP thank International Institute of Information Technology, Hyderabad for hospitality during the development of these ideas. AKP acknowledges the support of the J.C. Bose Fellowship from the Department of Science and Technology (DST), India under Grant No.~JCB/2018/000038 (2019–2024).

\bibliography{ref}

\begin{thebibliography}{67}%
\makeatletter
\providecommand \@ifxundefined [1]{%
 \@ifx{#1\undefined}
}%
\providecommand \@ifnum [1]{%
 \ifnum #1\expandafter \@firstoftwo
 \else \expandafter \@secondoftwo
 \fi
}%
\providecommand \@ifx [1]{%
 \ifx #1\expandafter \@firstoftwo
 \else \expandafter \@secondoftwo
 \fi
}%
\providecommand \natexlab [1]{#1}%
\providecommand \enquote  [1]{``#1''}%
\providecommand \bibnamefont  [1]{#1}%
\providecommand \bibfnamefont [1]{#1}%
\providecommand \citenamefont [1]{#1}%
\providecommand \href@noop [0]{\@secondoftwo}%
\providecommand \href [0]{\begingroup \@sanitize@url \@href}%
\providecommand \@href[1]{\@@startlink{#1}\@@href}%
\providecommand \@@href[1]{\endgroup#1\@@endlink}%
\providecommand \@sanitize@url [0]{\catcode `\\12\catcode `\$12\catcode `\&12\catcode `\#12\catcode `\^12\catcode `\_12\catcode `\%12\relax}%
\providecommand \@@startlink[1]{}%
\providecommand \@@endlink[0]{}%
\providecommand \url  [0]{\begingroup\@sanitize@url \@url }%
\providecommand \@url [1]{\endgroup\@href {#1}{\urlprefix }}%
\providecommand \urlprefix  [0]{URL }%
\providecommand \Eprint [0]{\href }%
\providecommand \doibase [0]{https://doi.org/}%
\providecommand \selectlanguage [0]{\@gobble}%
\providecommand \bibinfo  [0]{\@secondoftwo}%
\providecommand \bibfield  [0]{\@secondoftwo}%
\providecommand \translation [1]{[#1]}%
\providecommand \BibitemOpen [0]{}%
\providecommand \bibitemStop [0]{}%
\providecommand \bibitemNoStop [0]{.\EOS\space}%
\providecommand \EOS [0]{\spacefactor3000\relax}%
\providecommand \BibitemShut  [1]{\csname bibitem#1\endcsname}%
\let\auto@bib@innerbib\@empty
\bibitem [{\citenamefont {Mandelstam}\ and\ \citenamefont {Tamm}(1945)}]{Mandelstam1945}%
  \BibitemOpen
  \bibfield  {author} {\bibinfo {author} {\bibfnamefont {L.}~\bibnamefont {Mandelstam}}\ and\ \bibinfo {author} {\bibfnamefont {I.}~\bibnamefont {Tamm}},\ }\href {https://doi.org/10.1007/978-3-642-74626-0_8} {\bibfield  {journal} {\bibinfo  {journal} {J. Phys. (USSR)}\ }\textbf {\bibinfo {volume} {9}},\ \bibinfo {pages} {249} (\bibinfo {year} {1945})}\BibitemShut {NoStop}%
\bibitem [{\citenamefont {Anandan}\ and\ \citenamefont {Aharonov}(1990)}]{Anandan1990}%
  \BibitemOpen
  \bibfield  {author} {\bibinfo {author} {\bibfnamefont {J.}~\bibnamefont {Anandan}}\ and\ \bibinfo {author} {\bibfnamefont {Y.}~\bibnamefont {Aharonov}},\ }\href {https://doi.org/10.1103/PhysRevLett.65.1697} {\bibfield  {journal} {\bibinfo  {journal} {Physical Review Letters}\ }\textbf {\bibinfo {volume} {65}},\ \bibinfo {pages} {1697} (\bibinfo {year} {1990})}\BibitemShut {NoStop}%
\bibitem [{\citenamefont {Pati}(1991)}]{Pati_1991}%
  \BibitemOpen
  \bibfield  {author} {\bibinfo {author} {\bibfnamefont {A.~K.}\ \bibnamefont {Pati}},\ }\href {https://doi.org/https://doi.org/10.1016/0375-9601(91)90255-7} {\bibfield  {journal} {\bibinfo  {journal} {Physics Letters A}\ }\textbf {\bibinfo {volume} {159}},\ \bibinfo {pages} {105} (\bibinfo {year} {1991})}\BibitemShut {NoStop}%
\bibitem [{\citenamefont {Margolus}\ and\ \citenamefont {Levitin}(1998)}]{Margolus1998}%
  \BibitemOpen
  \bibfield  {author} {\bibinfo {author} {\bibfnamefont {N.}~\bibnamefont {Margolus}}\ and\ \bibinfo {author} {\bibfnamefont {L.~B.}\ \bibnamefont {Levitin}},\ }\href {https://doi.org/https://doi.org/10.1016/S0167-2789(98)00054-2} {\bibfield  {journal} {\bibinfo  {journal} {Physica D: Nonlinear Phenomena}\ }\textbf {\bibinfo {volume} {120}},\ \bibinfo {pages} {188} (\bibinfo {year} {1998})}\BibitemShut {NoStop}%
\bibitem [{\citenamefont {Bhattacharyya}(1983)}]{Bhattacharyya_1983}%
  \BibitemOpen
  \bibfield  {author} {\bibinfo {author} {\bibfnamefont {K.}~\bibnamefont {Bhattacharyya}},\ }\href {https://doi.org/10.1088/0305-4470/16/13/021} {\bibfield  {journal} {\bibinfo  {journal} {Journal of Physics A: Mathematical and General}\ }\textbf {\bibinfo {volume} {16}},\ \bibinfo {pages} {2993} (\bibinfo {year} {1983})}\BibitemShut {NoStop}%
\bibitem [{\citenamefont {Vaidman}(1992)}]{Vaidman1992}%
  \BibitemOpen
  \bibfield  {author} {\bibinfo {author} {\bibfnamefont {L.}~\bibnamefont {Vaidman}},\ }\href {https://doi.org/10.1119/1.16940} {\bibfield  {journal} {\bibinfo  {journal} {American Journal of Physics}\ }\textbf {\bibinfo {volume} {60}},\ \bibinfo {pages} {182} (\bibinfo {year} {1992})}\BibitemShut {NoStop}%
\bibitem [{\citenamefont {Uhlmann}(1992)}]{Uhlmann1992}%
  \BibitemOpen
  \bibfield  {author} {\bibinfo {author} {\bibfnamefont {A.}~\bibnamefont {Uhlmann}},\ }\href {https://doi.org/https://doi.org/10.1016/0375-9601(92)90555-Z} {\bibfield  {journal} {\bibinfo  {journal} {Physics Letters A}\ }\textbf {\bibinfo {volume} {161}},\ \bibinfo {pages} {329} (\bibinfo {year} {1992})}\BibitemShut {NoStop}%
\bibitem [{\citenamefont {Pfeifer}(1993)}]{Pfeifer1993}%
  \BibitemOpen
  \bibfield  {author} {\bibinfo {author} {\bibfnamefont {P.}~\bibnamefont {Pfeifer}},\ }\href {https://doi.org/10.1103/PhysRevLett.70.3365} {\bibfield  {journal} {\bibinfo  {journal} {Physical Review Letters}\ }\textbf {\bibinfo {volume} {70}},\ \bibinfo {pages} {3365} (\bibinfo {year} {1993})}\BibitemShut {NoStop}%
\bibitem [{\citenamefont {Levitin}\ and\ \citenamefont {Toffoli}(2009)}]{Levitin2009}%
  \BibitemOpen
  \bibfield  {author} {\bibinfo {author} {\bibfnamefont {L.~B.}\ \bibnamefont {Levitin}}\ and\ \bibinfo {author} {\bibfnamefont {T.}~\bibnamefont {Toffoli}},\ }\href {https://doi.org/10.1103/PhysRevLett.103.160502} {\bibfield  {journal} {\bibinfo  {journal} {Physical Review Letters}\ }\textbf {\bibinfo {volume} {103}},\ \bibinfo {pages} {160502} (\bibinfo {year} {2009})}\BibitemShut {NoStop}%
\bibitem [{\citenamefont {Deffner}\ and\ \citenamefont {Lutz}(2013{\natexlab{a}})}]{Deffner_2013}%
  \BibitemOpen
  \bibfield  {author} {\bibinfo {author} {\bibfnamefont {S.}~\bibnamefont {Deffner}}\ and\ \bibinfo {author} {\bibfnamefont {E.}~\bibnamefont {Lutz}},\ }\href {https://doi.org/10.1088/1751-8113/46/33/335302} {\bibfield  {journal} {\bibinfo  {journal} {Journal of Physics A: Mathematical and Theoretical}\ }\textbf {\bibinfo {volume} {46}},\ \bibinfo {pages} {335302} (\bibinfo {year} {2013}{\natexlab{a}})}\BibitemShut {NoStop}%
\bibitem [{\citenamefont {Mondal}\ and\ \citenamefont {Pati}(2016)}]{Mondal2016}%
  \BibitemOpen
  \bibfield  {author} {\bibinfo {author} {\bibfnamefont {D.}~\bibnamefont {Mondal}}\ and\ \bibinfo {author} {\bibfnamefont {A.~K.}\ \bibnamefont {Pati}},\ }\href {https://doi.org/https://doi.org/10.1016/j.physleta.2016.02.018} {\bibfield  {journal} {\bibinfo  {journal} {Physics Letters A}\ }\textbf {\bibinfo {volume} {380}},\ \bibinfo {pages} {1395} (\bibinfo {year} {2016})}\BibitemShut {NoStop}%
\bibitem [{\citenamefont {Mondal}\ \emph {et~al.}(2016)\citenamefont {Mondal}, \citenamefont {Datta},\ and\ \citenamefont {Sazim}}]{Mondal_2016}%
  \BibitemOpen
  \bibfield  {author} {\bibinfo {author} {\bibfnamefont {D.}~\bibnamefont {Mondal}}, \bibinfo {author} {\bibfnamefont {C.}~\bibnamefont {Datta}},\ and\ \bibinfo {author} {\bibfnamefont {S.}~\bibnamefont {Sazim}},\ }\href {https://doi.org/https://doi.org/10.1016/j.physleta.2015.12.015} {\bibfield  {journal} {\bibinfo  {journal} {Physics Letters A}\ }\textbf {\bibinfo {volume} {380}},\ \bibinfo {pages} {689} (\bibinfo {year} {2016})}\BibitemShut {NoStop}%
\bibitem [{\citenamefont {Campaioli}\ \emph {et~al.}(2018)\citenamefont {Campaioli}, \citenamefont {Pollock}, \citenamefont {Binder},\ and\ \citenamefont {Modi}}]{Campaioli2018}%
  \BibitemOpen
  \bibfield  {author} {\bibinfo {author} {\bibfnamefont {F.}~\bibnamefont {Campaioli}}, \bibinfo {author} {\bibfnamefont {F.~A.}\ \bibnamefont {Pollock}}, \bibinfo {author} {\bibfnamefont {F.~C.}\ \bibnamefont {Binder}},\ and\ \bibinfo {author} {\bibfnamefont {K.}~\bibnamefont {Modi}},\ }\href {https://doi.org/10.1103/PhysRevLett.120.060409} {\bibfield  {journal} {\bibinfo  {journal} {Physical Review Letters}\ }\textbf {\bibinfo {volume} {120}},\ \bibinfo {pages} {060409} (\bibinfo {year} {2018})}\BibitemShut {NoStop}%
\bibitem [{\citenamefont {Shao}\ \emph {et~al.}(2020)\citenamefont {Shao}, \citenamefont {Liu}, \citenamefont {Zhang}, \citenamefont {Yuan},\ and\ \citenamefont {Liu}}]{Shao2020}%
  \BibitemOpen
  \bibfield  {author} {\bibinfo {author} {\bibfnamefont {Y.}~\bibnamefont {Shao}}, \bibinfo {author} {\bibfnamefont {B.}~\bibnamefont {Liu}}, \bibinfo {author} {\bibfnamefont {M.}~\bibnamefont {Zhang}}, \bibinfo {author} {\bibfnamefont {H.}~\bibnamefont {Yuan}},\ and\ \bibinfo {author} {\bibfnamefont {J.}~\bibnamefont {Liu}},\ }\href {https://doi.org/10.1103/PhysRevResearch.2.023299} {\bibfield  {journal} {\bibinfo  {journal} {Physical Review Research}\ }\textbf {\bibinfo {volume} {2}},\ \bibinfo {pages} {023299} (\bibinfo {year} {2020})}\BibitemShut {NoStop}%
\bibitem [{\citenamefont {Ness}\ \emph {et~al.}(2021)\citenamefont {Ness}, \citenamefont {Lam}, \citenamefont {Alt}, \citenamefont {Meschede}, \citenamefont {Sagi},\ and\ \citenamefont {Alberti}}]{Ness2021}%
  \BibitemOpen
  \bibfield  {author} {\bibinfo {author} {\bibfnamefont {G.}~\bibnamefont {Ness}}, \bibinfo {author} {\bibfnamefont {M.~R.}\ \bibnamefont {Lam}}, \bibinfo {author} {\bibfnamefont {W.}~\bibnamefont {Alt}}, \bibinfo {author} {\bibfnamefont {D.}~\bibnamefont {Meschede}}, \bibinfo {author} {\bibfnamefont {Y.}~\bibnamefont {Sagi}},\ and\ \bibinfo {author} {\bibfnamefont {A.}~\bibnamefont {Alberti}},\ }\href {https://doi.org/10.1126/sciadv.abj9119} {\bibfield  {journal} {\bibinfo  {journal} {Science Advances}\ }\textbf {\bibinfo {volume} {7}},\ \bibinfo {pages} {eabj9119} (\bibinfo {year} {2021})}\BibitemShut {NoStop}%
\bibitem [{\citenamefont {Ness}\ \emph {et~al.}(2022)\citenamefont {Ness}, \citenamefont {Alberti},\ and\ \citenamefont {Sagi}}]{Ness2022}%
  \BibitemOpen
  \bibfield  {author} {\bibinfo {author} {\bibfnamefont {G.}~\bibnamefont {Ness}}, \bibinfo {author} {\bibfnamefont {A.}~\bibnamefont {Alberti}},\ and\ \bibinfo {author} {\bibfnamefont {Y.}~\bibnamefont {Sagi}},\ }\href {https://doi.org/10.1103/PhysRevLett.129.140403} {\bibfield  {journal} {\bibinfo  {journal} {Physical Review Letters}\ }\textbf {\bibinfo {volume} {129}},\ \bibinfo {pages} {140403} (\bibinfo {year} {2022})}\BibitemShut {NoStop}%
\bibitem [{\citenamefont {Bagchi}\ \emph {et~al.}(2022)\citenamefont {Bagchi}, \citenamefont {Srivastav},\ and\ \citenamefont {Pati}}]{Bagchi2022}%
  \BibitemOpen
  \bibfield  {author} {\bibinfo {author} {\bibfnamefont {S.}~\bibnamefont {Bagchi}}, \bibinfo {author} {\bibfnamefont {A.}~\bibnamefont {Srivastav}},\ and\ \bibinfo {author} {\bibfnamefont {A.~K.}\ \bibnamefont {Pati}},\ }\href {https://doi.org/10.48550/arXiv.2211.14561} {\bibfield  {journal} {\bibinfo  {journal} {arXiv:2211.14561}\ } (\bibinfo {year} {2022})}\BibitemShut {NoStop}%
\bibitem [{\citenamefont {Bender}\ \emph {et~al.}(2007)\citenamefont {Bender}, \citenamefont {Brody}, \citenamefont {Jones},\ and\ \citenamefont {Meister}}]{Bender2007}%
  \BibitemOpen
  \bibfield  {author} {\bibinfo {author} {\bibfnamefont {C.~M.}\ \bibnamefont {Bender}}, \bibinfo {author} {\bibfnamefont {D.~C.}\ \bibnamefont {Brody}}, \bibinfo {author} {\bibfnamefont {H.~F.}\ \bibnamefont {Jones}},\ and\ \bibinfo {author} {\bibfnamefont {B.~K.}\ \bibnamefont {Meister}},\ }\href {https://doi.org/10.1103/PhysRevLett.98.040403} {\bibfield  {journal} {\bibinfo  {journal} {Physical Review Letters}\ }\textbf {\bibinfo {volume} {98}},\ \bibinfo {pages} {040403} (\bibinfo {year} {2007})}\BibitemShut {NoStop}%
\bibitem [{\citenamefont {Uzdin}\ \emph {et~al.}(2012)\citenamefont {Uzdin}, \citenamefont {Günther}, \citenamefont {Rahav},\ and\ \citenamefont {Moiseyev}}]{Uzdin_2012}%
  \BibitemOpen
  \bibfield  {author} {\bibinfo {author} {\bibfnamefont {R.}~\bibnamefont {Uzdin}}, \bibinfo {author} {\bibfnamefont {U.}~\bibnamefont {Günther}}, \bibinfo {author} {\bibfnamefont {S.}~\bibnamefont {Rahav}},\ and\ \bibinfo {author} {\bibfnamefont {N.}~\bibnamefont {Moiseyev}},\ }\href {https://doi.org/10.1088/1751-8113/45/41/415304} {\bibfield  {journal} {\bibinfo  {journal} {Journal of Physics A: Mathematical and Theoretical}\ }\textbf {\bibinfo {volume} {45}},\ \bibinfo {pages} {415304} (\bibinfo {year} {2012})}\BibitemShut {NoStop}%
\bibitem [{\citenamefont {del Campo}\ \emph {et~al.}(2013)\citenamefont {del Campo}, \citenamefont {Egusquiza}, \citenamefont {Plenio},\ and\ \citenamefont {Huelga}}]{delCampo2013}%
  \BibitemOpen
  \bibfield  {author} {\bibinfo {author} {\bibfnamefont {A.}~\bibnamefont {del Campo}}, \bibinfo {author} {\bibfnamefont {I.~L.}\ \bibnamefont {Egusquiza}}, \bibinfo {author} {\bibfnamefont {M.~B.}\ \bibnamefont {Plenio}},\ and\ \bibinfo {author} {\bibfnamefont {S.~F.}\ \bibnamefont {Huelga}},\ }\href {https://doi.org/10.1103/PhysRevLett.110.050403} {\bibfield  {journal} {\bibinfo  {journal} {Physical Review Letters}\ }\textbf {\bibinfo {volume} {110}},\ \bibinfo {pages} {050403} (\bibinfo {year} {2013})}\BibitemShut {NoStop}%
\bibitem [{\citenamefont {Taddei}\ \emph {et~al.}(2013)\citenamefont {Taddei}, \citenamefont {Escher}, \citenamefont {Davidovich},\ and\ \citenamefont {de~Matos~Filho}}]{Taddei2013}%
  \BibitemOpen
  \bibfield  {author} {\bibinfo {author} {\bibfnamefont {M.~M.}\ \bibnamefont {Taddei}}, \bibinfo {author} {\bibfnamefont {B.~M.}\ \bibnamefont {Escher}}, \bibinfo {author} {\bibfnamefont {L.}~\bibnamefont {Davidovich}},\ and\ \bibinfo {author} {\bibfnamefont {R.~L.}\ \bibnamefont {de~Matos~Filho}},\ }\href {https://doi.org/10.1103/PhysRevLett.110.050402} {\bibfield  {journal} {\bibinfo  {journal} {Physical Review Letters}\ }\textbf {\bibinfo {volume} {110}},\ \bibinfo {pages} {050402} (\bibinfo {year} {2013})}\BibitemShut {NoStop}%
\bibitem [{\citenamefont {Deffner}\ and\ \citenamefont {Lutz}(2013{\natexlab{b}})}]{Deffner2013}%
  \BibitemOpen
  \bibfield  {author} {\bibinfo {author} {\bibfnamefont {S.}~\bibnamefont {Deffner}}\ and\ \bibinfo {author} {\bibfnamefont {E.}~\bibnamefont {Lutz}},\ }\href {https://doi.org/10.1103/PhysRevLett.111.010402} {\bibfield  {journal} {\bibinfo  {journal} {Physical Review Letters}\ }\textbf {\bibinfo {volume} {111}},\ \bibinfo {pages} {010402} (\bibinfo {year} {2013}{\natexlab{b}})}\BibitemShut {NoStop}%
\bibitem [{\citenamefont {Pires}\ \emph {et~al.}(2016)\citenamefont {Pires}, \citenamefont {Cianciaruso}, \citenamefont {C\'eleri}, \citenamefont {Adesso},\ and\ \citenamefont {Soares-Pinto}}]{Pires2016}%
  \BibitemOpen
  \bibfield  {author} {\bibinfo {author} {\bibfnamefont {D.~P.}\ \bibnamefont {Pires}}, \bibinfo {author} {\bibfnamefont {M.}~\bibnamefont {Cianciaruso}}, \bibinfo {author} {\bibfnamefont {L.~C.}\ \bibnamefont {C\'eleri}}, \bibinfo {author} {\bibfnamefont {G.}~\bibnamefont {Adesso}},\ and\ \bibinfo {author} {\bibfnamefont {D.~O.}\ \bibnamefont {Soares-Pinto}},\ }\href {https://doi.org/10.1103/PhysRevX.6.021031} {\bibfield  {journal} {\bibinfo  {journal} {Physical Review X}\ }\textbf {\bibinfo {volume} {6}},\ \bibinfo {pages} {021031} (\bibinfo {year} {2016})}\BibitemShut {NoStop}%
\bibitem [{\citenamefont {Campaioli}\ \emph {et~al.}(2019)\citenamefont {Campaioli}, \citenamefont {Pollock},\ and\ \citenamefont {Modi}}]{Campaioli2019}%
  \BibitemOpen
  \bibfield  {author} {\bibinfo {author} {\bibfnamefont {F.}~\bibnamefont {Campaioli}}, \bibinfo {author} {\bibfnamefont {F.~A.}\ \bibnamefont {Pollock}},\ and\ \bibinfo {author} {\bibfnamefont {K.}~\bibnamefont {Modi}},\ }\href {https://doi.org/10.22331/q-2019-08-05-168} {\bibfield  {journal} {\bibinfo  {journal} {{Quantum}}\ }\textbf {\bibinfo {volume} {3}},\ \bibinfo {pages} {168} (\bibinfo {year} {2019})}\BibitemShut {NoStop}%
\bibitem [{\citenamefont {Brody}\ and\ \citenamefont {Longstaff}(2019)}]{Brody2019}%
  \BibitemOpen
  \bibfield  {author} {\bibinfo {author} {\bibfnamefont {D.~C.}\ \bibnamefont {Brody}}\ and\ \bibinfo {author} {\bibfnamefont {B.}~\bibnamefont {Longstaff}},\ }\href {https://doi.org/10.1103/PhysRevResearch.1.033127} {\bibfield  {journal} {\bibinfo  {journal} {Physical Review Research}\ }\textbf {\bibinfo {volume} {1}},\ \bibinfo {pages} {033127} (\bibinfo {year} {2019})}\BibitemShut {NoStop}%
\bibitem [{\citenamefont {Das}\ \emph {et~al.}(2021)\citenamefont {Das}, \citenamefont {Bera}, \citenamefont {Chakraborty},\ and\ \citenamefont {Chru\ifmmode \acute{s}\else \'{s}\fi{}ci\ifmmode~\acute{n}\else \'{n}\fi{}ski}}]{Das2021}%
  \BibitemOpen
  \bibfield  {author} {\bibinfo {author} {\bibfnamefont {A.}~\bibnamefont {Das}}, \bibinfo {author} {\bibfnamefont {A.}~\bibnamefont {Bera}}, \bibinfo {author} {\bibfnamefont {S.}~\bibnamefont {Chakraborty}},\ and\ \bibinfo {author} {\bibfnamefont {D.}~\bibnamefont {Chru\ifmmode \acute{s}\else \'{s}\fi{}ci\ifmmode~\acute{n}\else \'{n}\fi{}ski}},\ }\href {https://doi.org/10.1103/PhysRevA.104.042202} {\bibfield  {journal} {\bibinfo  {journal} {Physical Review A}\ }\textbf {\bibinfo {volume} {104}},\ \bibinfo {pages} {042202} (\bibinfo {year} {2021})}\BibitemShut {NoStop}%
\bibitem [{\citenamefont {Impens}\ \emph {et~al.}(2021)\citenamefont {Impens}, \citenamefont {D'Angelis}, \citenamefont {Pinheiro},\ and\ \citenamefont {Gu\'ery-Odelin}}]{Impens2021}%
  \BibitemOpen
  \bibfield  {author} {\bibinfo {author} {\bibfnamefont {F.}~\bibnamefont {Impens}}, \bibinfo {author} {\bibfnamefont {F.~M.}\ \bibnamefont {D'Angelis}}, \bibinfo {author} {\bibfnamefont {F.~A.}\ \bibnamefont {Pinheiro}},\ and\ \bibinfo {author} {\bibfnamefont {D.}~\bibnamefont {Gu\'ery-Odelin}},\ }\href {https://doi.org/10.1103/PhysRevA.104.052620} {\bibfield  {journal} {\bibinfo  {journal} {Physical Review A}\ }\textbf {\bibinfo {volume} {104}},\ \bibinfo {pages} {052620} (\bibinfo {year} {2021})}\BibitemShut {NoStop}%
\bibitem [{\citenamefont {Lan}\ \emph {et~al.}(2022)\citenamefont {Lan}, \citenamefont {Xie},\ and\ \citenamefont {Cai}}]{Lan_2022}%
  \BibitemOpen
  \bibfield  {author} {\bibinfo {author} {\bibfnamefont {K.}~\bibnamefont {Lan}}, \bibinfo {author} {\bibfnamefont {S.}~\bibnamefont {Xie}},\ and\ \bibinfo {author} {\bibfnamefont {X.}~\bibnamefont {Cai}},\ }\href {https://doi.org/10.1088/1367-2630/ac696b} {\bibfield  {journal} {\bibinfo  {journal} {New Journal of Physics}\ }\textbf {\bibinfo {volume} {24}},\ \bibinfo {pages} {055003} (\bibinfo {year} {2022})}\BibitemShut {NoStop}%
\bibitem [{\citenamefont {Nakajima}\ and\ \citenamefont {Utsumi}(2022)}]{Nakajima_2022}%
  \BibitemOpen
  \bibfield  {author} {\bibinfo {author} {\bibfnamefont {S.}~\bibnamefont {Nakajima}}\ and\ \bibinfo {author} {\bibfnamefont {Y.}~\bibnamefont {Utsumi}},\ }\href {https://doi.org/10.1088/1367-2630/ac8eca} {\bibfield  {journal} {\bibinfo  {journal} {New Journal of Physics}\ }\textbf {\bibinfo {volume} {24}},\ \bibinfo {pages} {095004} (\bibinfo {year} {2022})}\BibitemShut {NoStop}%
\bibitem [{\citenamefont {Thakuria}\ \emph {et~al.}(2023)\citenamefont {Thakuria}, \citenamefont {Srivastav}, \citenamefont {Mohan}, \citenamefont {Kumari},\ and\ \citenamefont {Pati}}]{Thakuria_2024}%
  \BibitemOpen
  \bibfield  {author} {\bibinfo {author} {\bibfnamefont {D.}~\bibnamefont {Thakuria}}, \bibinfo {author} {\bibfnamefont {A.}~\bibnamefont {Srivastav}}, \bibinfo {author} {\bibfnamefont {B.}~\bibnamefont {Mohan}}, \bibinfo {author} {\bibfnamefont {A.}~\bibnamefont {Kumari}},\ and\ \bibinfo {author} {\bibfnamefont {A.~K.}\ \bibnamefont {Pati}},\ }\href {https://doi.org/10.1088/1751-8121/ad15ad} {\bibfield  {journal} {\bibinfo  {journal} {Journal of Physics A: Mathematical and Theoretical}\ }\textbf {\bibinfo {volume} {57}},\ \bibinfo {pages} {025302} (\bibinfo {year} {2023})}\BibitemShut {NoStop}%
\bibitem [{\citenamefont {Giovannetti}\ \emph {et~al.}(2004)\citenamefont {Giovannetti}, \citenamefont {Lloyd},\ and\ \citenamefont {Maccone}}]{Giovannetti_2004}%
  \BibitemOpen
  \bibfield  {author} {\bibinfo {author} {\bibfnamefont {V.}~\bibnamefont {Giovannetti}}, \bibinfo {author} {\bibfnamefont {S.}~\bibnamefont {Lloyd}},\ and\ \bibinfo {author} {\bibfnamefont {L.}~\bibnamefont {Maccone}},\ }\href {https://doi.org/10.1088/1464-4266/6/8/028} {\bibfield  {journal} {\bibinfo  {journal} {Journal of Optics B: Quantum and Semiclassical Optics}\ }\textbf {\bibinfo {volume} {6}},\ \bibinfo {pages} {S807} (\bibinfo {year} {2004})}\BibitemShut {NoStop}%
\bibitem [{\citenamefont {Batle}\ \emph {et~al.}(2005)\citenamefont {Batle}, \citenamefont {Casas}, \citenamefont {Plastino},\ and\ \citenamefont {Plastino}}]{Batle2005}%
  \BibitemOpen
  \bibfield  {author} {\bibinfo {author} {\bibfnamefont {J.}~\bibnamefont {Batle}}, \bibinfo {author} {\bibfnamefont {M.}~\bibnamefont {Casas}}, \bibinfo {author} {\bibfnamefont {A.}~\bibnamefont {Plastino}},\ and\ \bibinfo {author} {\bibfnamefont {A.~R.}\ \bibnamefont {Plastino}},\ }\href {https://doi.org/10.1103/PhysRevA.72.032337} {\bibfield  {journal} {\bibinfo  {journal} {Physical Reviwe A}\ }\textbf {\bibinfo {volume} {72}},\ \bibinfo {pages} {032337} (\bibinfo {year} {2005})}\BibitemShut {NoStop}%
\bibitem [{\citenamefont {Zander}\ \emph {et~al.}(2007)\citenamefont {Zander}, \citenamefont {Plastino}, \citenamefont {Plastino},\ and\ \citenamefont {Casas}}]{Zander_2007}%
  \BibitemOpen
  \bibfield  {author} {\bibinfo {author} {\bibfnamefont {C.}~\bibnamefont {Zander}}, \bibinfo {author} {\bibfnamefont {A.~R.}\ \bibnamefont {Plastino}}, \bibinfo {author} {\bibfnamefont {A.}~\bibnamefont {Plastino}},\ and\ \bibinfo {author} {\bibfnamefont {M.}~\bibnamefont {Casas}},\ }\href {https://doi.org/10.1088/1751-8113/40/11/020} {\bibfield  {journal} {\bibinfo  {journal} {Journal of Physics A: Mathematical and Theoretical}\ }\textbf {\bibinfo {volume} {40}},\ \bibinfo {pages} {2861} (\bibinfo {year} {2007})}\BibitemShut {NoStop}%
\bibitem [{\citenamefont {Borras}\ \emph {et~al.}(2007)\citenamefont {Borras}, \citenamefont {Zander}, \citenamefont {Plastino}, \citenamefont {Casas},\ and\ \citenamefont {Plastino}}]{Borras_2008}%
  \BibitemOpen
  \bibfield  {author} {\bibinfo {author} {\bibfnamefont {A.}~\bibnamefont {Borras}}, \bibinfo {author} {\bibfnamefont {C.}~\bibnamefont {Zander}}, \bibinfo {author} {\bibfnamefont {A.~R.}\ \bibnamefont {Plastino}}, \bibinfo {author} {\bibfnamefont {M.}~\bibnamefont {Casas}},\ and\ \bibinfo {author} {\bibfnamefont {A.}~\bibnamefont {Plastino}},\ }\href {https://doi.org/10.1209/0295-5075/81/30007} {\bibfield  {journal} {\bibinfo  {journal} {Europhysics Letters}\ }\textbf {\bibinfo {volume} {81}},\ \bibinfo {pages} {30007} (\bibinfo {year} {2007})}\BibitemShut {NoStop}%
\bibitem [{\citenamefont {Brouzos}\ \emph {et~al.}(2015)\citenamefont {Brouzos}, \citenamefont {Streltsov}, \citenamefont {Negretti}, \citenamefont {Said}, \citenamefont {Caneva}, \citenamefont {Montangero},\ and\ \citenamefont {Calarco}}]{Brouzos2015}%
  \BibitemOpen
  \bibfield  {author} {\bibinfo {author} {\bibfnamefont {I.}~\bibnamefont {Brouzos}}, \bibinfo {author} {\bibfnamefont {A.~I.}\ \bibnamefont {Streltsov}}, \bibinfo {author} {\bibfnamefont {A.}~\bibnamefont {Negretti}}, \bibinfo {author} {\bibfnamefont {R.~S.}\ \bibnamefont {Said}}, \bibinfo {author} {\bibfnamefont {T.}~\bibnamefont {Caneva}}, \bibinfo {author} {\bibfnamefont {S.}~\bibnamefont {Montangero}},\ and\ \bibinfo {author} {\bibfnamefont {T.}~\bibnamefont {Calarco}},\ }\href {https://doi.org/10.1103/PhysRevA.92.062110} {\bibfield  {journal} {\bibinfo  {journal} {Physical Review A}\ }\textbf {\bibinfo {volume} {92}},\ \bibinfo {pages} {062110} (\bibinfo {year} {2015})}\BibitemShut {NoStop}%
\bibitem [{\citenamefont {Bukov}\ \emph {et~al.}(2019)\citenamefont {Bukov}, \citenamefont {Sels},\ and\ \citenamefont {Polkovnikov}}]{Bukov2019}%
  \BibitemOpen
  \bibfield  {author} {\bibinfo {author} {\bibfnamefont {M.}~\bibnamefont {Bukov}}, \bibinfo {author} {\bibfnamefont {D.}~\bibnamefont {Sels}},\ and\ \bibinfo {author} {\bibfnamefont {A.}~\bibnamefont {Polkovnikov}},\ }\href {https://doi.org/10.1103/PhysRevX.9.011034} {\bibfield  {journal} {\bibinfo  {journal} {Physical Review X}\ }\textbf {\bibinfo {volume} {9}},\ \bibinfo {pages} {011034} (\bibinfo {year} {2019})}\BibitemShut {NoStop}%
\bibitem [{\citenamefont {Van~Vu}\ and\ \citenamefont {Saito}(2023)}]{Van2023}%
  \BibitemOpen
  \bibfield  {author} {\bibinfo {author} {\bibfnamefont {T.}~\bibnamefont {Van~Vu}}\ and\ \bibinfo {author} {\bibfnamefont {K.}~\bibnamefont {Saito}},\ }\href {https://doi.org/10.1103/PhysRevLett.130.010402} {\bibfield  {journal} {\bibinfo  {journal} {Physical Review Letters}\ }\textbf {\bibinfo {volume} {130}},\ \bibinfo {pages} {010402} (\bibinfo {year} {2023})}\BibitemShut {NoStop}%
\bibitem [{\citenamefont {Pati}\ \emph {et~al.}(2023)\citenamefont {Pati}, \citenamefont {Mohan}, \citenamefont {Sahil},\ and\ \citenamefont {Braunstein}}]{Pati2023}%
  \BibitemOpen
  \bibfield  {author} {\bibinfo {author} {\bibfnamefont {A.~K.}\ \bibnamefont {Pati}}, \bibinfo {author} {\bibfnamefont {B.}~\bibnamefont {Mohan}}, \bibinfo {author} {\bibnamefont {Sahil}},\ and\ \bibinfo {author} {\bibfnamefont {S.~L.}\ \bibnamefont {Braunstein}},\ }\href {https://doi.org/10.48550/arXiv.2305.03839} {\bibfield  {journal} {\bibinfo  {journal} {arXiv:2305.03839}\ } (\bibinfo {year} {2023})}\BibitemShut {NoStop}%
\bibitem [{\citenamefont {Garc\'{\i}a-Pintos}\ \emph {et~al.}(2022)\citenamefont {Garc\'{\i}a-Pintos}, \citenamefont {Nicholson}, \citenamefont {Green}, \citenamefont {del Campo},\ and\ \citenamefont {Gorshkov}}]{Pintos2022}%
  \BibitemOpen
  \bibfield  {author} {\bibinfo {author} {\bibfnamefont {L.~P.}\ \bibnamefont {Garc\'{\i}a-Pintos}}, \bibinfo {author} {\bibfnamefont {S.~B.}\ \bibnamefont {Nicholson}}, \bibinfo {author} {\bibfnamefont {J.~R.}\ \bibnamefont {Green}}, \bibinfo {author} {\bibfnamefont {A.}~\bibnamefont {del Campo}},\ and\ \bibinfo {author} {\bibfnamefont {A.~V.}\ \bibnamefont {Gorshkov}},\ }\href {https://doi.org/10.1103/PhysRevX.12.011038} {\bibfield  {journal} {\bibinfo  {journal} {Physical Review X}\ }\textbf {\bibinfo {volume} {12}},\ \bibinfo {pages} {011038} (\bibinfo {year} {2022})}\BibitemShut {NoStop}%
\bibitem [{\citenamefont {Hamazaki}(2022)}]{Hamazaki2022}%
  \BibitemOpen
  \bibfield  {author} {\bibinfo {author} {\bibfnamefont {R.}~\bibnamefont {Hamazaki}},\ }\href {https://doi.org/10.1103/PRXQuantum.3.020319} {\bibfield  {journal} {\bibinfo  {journal} {PRX Quantum}\ }\textbf {\bibinfo {volume} {3}},\ \bibinfo {pages} {020319} (\bibinfo {year} {2022})}\BibitemShut {NoStop}%
\bibitem [{\citenamefont {Mohan}\ and\ \citenamefont {Pati}(2022)}]{Mohan2022}%
  \BibitemOpen
  \bibfield  {author} {\bibinfo {author} {\bibfnamefont {B.}~\bibnamefont {Mohan}}\ and\ \bibinfo {author} {\bibfnamefont {A.~K.}\ \bibnamefont {Pati}},\ }\href {https://doi.org/10.1103/PhysRevA.106.042436} {\bibfield  {journal} {\bibinfo  {journal} {Physical Review A}\ }\textbf {\bibinfo {volume} {106}},\ \bibinfo {pages} {042436} (\bibinfo {year} {2022})}\BibitemShut {NoStop}%
\bibitem [{\citenamefont {H{\"{o}}rnedal}\ \emph {et~al.}(2023)\citenamefont {H{\"{o}}rnedal}, \citenamefont {Carabba}, \citenamefont {Takahashi},\ and\ \citenamefont {del Campo}}]{Hornedal2023}%
  \BibitemOpen
  \bibfield  {author} {\bibinfo {author} {\bibfnamefont {N.}~\bibnamefont {H{\"{o}}rnedal}}, \bibinfo {author} {\bibfnamefont {N.}~\bibnamefont {Carabba}}, \bibinfo {author} {\bibfnamefont {K.}~\bibnamefont {Takahashi}},\ and\ \bibinfo {author} {\bibfnamefont {A.}~\bibnamefont {del Campo}},\ }\href {https://doi.org/10.22331/q-2023-07-11-1055} {\bibfield  {journal} {\bibinfo  {journal} {{Quantum}}\ }\textbf {\bibinfo {volume} {7}},\ \bibinfo {pages} {1055} (\bibinfo {year} {2023})}\BibitemShut {NoStop}%
\bibitem [{\citenamefont {Mohan}\ \emph {et~al.}(2022)\citenamefont {Mohan}, \citenamefont {Das},\ and\ \citenamefont {Pati}}]{Mohan_2022}%
  \BibitemOpen
  \bibfield  {author} {\bibinfo {author} {\bibfnamefont {B.}~\bibnamefont {Mohan}}, \bibinfo {author} {\bibfnamefont {S.}~\bibnamefont {Das}},\ and\ \bibinfo {author} {\bibfnamefont {A.~K.}\ \bibnamefont {Pati}},\ }\href {https://doi.org/10.1088/1367-2630/ac753c} {\bibfield  {journal} {\bibinfo  {journal} {New Journal of Physics}\ }\textbf {\bibinfo {volume} {24}},\ \bibinfo {pages} {065003} (\bibinfo {year} {2022})}\BibitemShut {NoStop}%
\bibitem [{\citenamefont {Pandey}\ \emph {et~al.}(2023)\citenamefont {Pandey}, \citenamefont {Bhowmick}, \citenamefont {Mohan}, \citenamefont {Sohail},\ and\ \citenamefont {Sen}}]{Pandey2023}%
  \BibitemOpen
  \bibfield  {author} {\bibinfo {author} {\bibfnamefont {V.}~\bibnamefont {Pandey}}, \bibinfo {author} {\bibfnamefont {S.}~\bibnamefont {Bhowmick}}, \bibinfo {author} {\bibfnamefont {B.}~\bibnamefont {Mohan}}, \bibinfo {author} {\bibnamefont {Sohail}},\ and\ \bibinfo {author} {\bibfnamefont {U.}~\bibnamefont {Sen}},\ }\href {https://doi.org/10.48550/arXiv.2303.07415} {\bibfield  {journal} {\bibinfo  {journal} {arXiv:2303.07415}\ } (\bibinfo {year} {2023})}\BibitemShut {NoStop}%
\bibitem [{\citenamefont {Maccone}\ and\ \citenamefont {Pati}(2014)}]{Maccone2014}%
  \BibitemOpen
  \bibfield  {author} {\bibinfo {author} {\bibfnamefont {L.}~\bibnamefont {Maccone}}\ and\ \bibinfo {author} {\bibfnamefont {A.~K.}\ \bibnamefont {Pati}},\ }\href {https://doi.org/10.1103/PhysRevLett.113.260401} {\bibfield  {journal} {\bibinfo  {journal} {Phys. Rev. Lett.}\ }\textbf {\bibinfo {volume} {113}},\ \bibinfo {pages} {260401} (\bibinfo {year} {2014})}\BibitemShut {NoStop}%
\bibitem [{\citenamefont {Thakuria}\ and\ \citenamefont {Pati}(2022)}]{thakuria2022}%
  \BibitemOpen
  \bibfield  {author} {\bibinfo {author} {\bibfnamefont {D.}~\bibnamefont {Thakuria}}\ and\ \bibinfo {author} {\bibfnamefont {A.~K.}\ \bibnamefont {Pati}},\ }\href@noop {} {} (\bibinfo {year} {2022}),\ \Eprint {https://arxiv.org/abs/2208.05469} {arXiv:2208.05469} \BibitemShut {NoStop}%
\bibitem [{\citenamefont {Shrimali}\ \emph {et~al.}(2024)\citenamefont {Shrimali}, \citenamefont {Panda},\ and\ \citenamefont {Pati}}]{Divyansh2024}%
  \BibitemOpen
  \bibfield  {author} {\bibinfo {author} {\bibfnamefont {D.}~\bibnamefont {Shrimali}}, \bibinfo {author} {\bibfnamefont {B.}~\bibnamefont {Panda}},\ and\ \bibinfo {author} {\bibfnamefont {A.}~\bibnamefont {Pati}},\ }\href@noop {} {} (\bibinfo {year} {2024}),\ \Eprint {https://arxiv.org/abs/2404.03247} {arXiv:2404.03247} \BibitemShut {NoStop}%
\bibitem [{\citenamefont {Misra}\ and\ \citenamefont {Sudarshan}(1977)}]{Misra1977}%
  \BibitemOpen
  \bibfield  {author} {\bibinfo {author} {\bibfnamefont {B.}~\bibnamefont {Misra}}\ and\ \bibinfo {author} {\bibfnamefont {E.~C.~G.}\ \bibnamefont {Sudarshan}},\ }\href {https://doi.org/10.1063/1.523304} {\bibfield  {journal} {\bibinfo  {journal} {Journal of Mathematical Physics}\ }\textbf {\bibinfo {volume} {18}},\ \bibinfo {pages} {756} (\bibinfo {year} {1977})}\BibitemShut {NoStop}%
\bibitem [{\citenamefont {Itano}\ \emph {et~al.}(1990)\citenamefont {Itano}, \citenamefont {Heinzen}, \citenamefont {Bollinger},\ and\ \citenamefont {Wineland}}]{Itano1990}%
  \BibitemOpen
  \bibfield  {author} {\bibinfo {author} {\bibfnamefont {W.~M.}\ \bibnamefont {Itano}}, \bibinfo {author} {\bibfnamefont {D.~J.}\ \bibnamefont {Heinzen}}, \bibinfo {author} {\bibfnamefont {J.~J.}\ \bibnamefont {Bollinger}},\ and\ \bibinfo {author} {\bibfnamefont {D.~J.}\ \bibnamefont {Wineland}},\ }\href {https://doi.org/10.1103/PhysRevA.41.2295} {\bibfield  {journal} {\bibinfo  {journal} {Physical Review A}\ }\textbf {\bibinfo {volume} {41}},\ \bibinfo {pages} {2295} (\bibinfo {year} {1990})}\BibitemShut {NoStop}%
\bibitem [{\citenamefont {Braginsky}\ \emph {et~al.}(1995)\citenamefont {Braginsky}, \citenamefont {Khalili},\ and\ \citenamefont {Kulaga}}]{Braginsky1995}%
  \BibitemOpen
  \bibfield  {author} {\bibinfo {author} {\bibfnamefont {V.}~\bibnamefont {Braginsky}}, \bibinfo {author} {\bibfnamefont {F.}~\bibnamefont {Khalili}},\ and\ \bibinfo {author} {\bibfnamefont {A.}~\bibnamefont {Kulaga}},\ }\href {https://doi.org/https://doi.org/10.1016/0375-9601(95)00279-C} {\bibfield  {journal} {\bibinfo  {journal} {Physics Letters A}\ }\textbf {\bibinfo {volume} {202}},\ \bibinfo {pages} {1} (\bibinfo {year} {1995})}\BibitemShut {NoStop}%
\bibitem [{\citenamefont {Pati}\ and\ \citenamefont {Lawande}(1996)}]{Pati_1996}%
  \BibitemOpen
  \bibfield  {author} {\bibinfo {author} {\bibfnamefont {A.~K.}\ \bibnamefont {Pati}}\ and\ \bibinfo {author} {\bibfnamefont {S.~V.}\ \bibnamefont {Lawande}},\ }\href {https://doi.org/https://doi.org/10.1016/S0375-9601(96)00699-8} {\bibfield  {journal} {\bibinfo  {journal} {Physics Letters A}\ }\textbf {\bibinfo {volume} {223}},\ \bibinfo {pages} {233} (\bibinfo {year} {1996})}\BibitemShut {NoStop}%
\bibitem [{\citenamefont {Beige}\ and\ \citenamefont {Hegerfeldt}(1997)}]{Beige_1997}%
  \BibitemOpen
  \bibfield  {author} {\bibinfo {author} {\bibfnamefont {A.}~\bibnamefont {Beige}}\ and\ \bibinfo {author} {\bibfnamefont {G.~C.}\ \bibnamefont {Hegerfeldt}},\ }\href {https://doi.org/10.1088/0305-4470/30/4/031} {\bibfield  {journal} {\bibinfo  {journal} {Journal of Physics A: Mathematical and General}\ }\textbf {\bibinfo {volume} {30}},\ \bibinfo {pages} {1323} (\bibinfo {year} {1997})}\BibitemShut {NoStop}%
\bibitem [{\citenamefont {Sj\"oqvist}\ and\ \citenamefont {Carlsen}(1997)}]{Erik1997}%
  \BibitemOpen
  \bibfield  {author} {\bibinfo {author} {\bibfnamefont {E.}~\bibnamefont {Sj\"oqvist}}\ and\ \bibinfo {author} {\bibfnamefont {H.}~\bibnamefont {Carlsen}},\ }\href {https://doi.org/10.1103/PhysRevA.56.1638} {\bibfield  {journal} {\bibinfo  {journal} {Physical Review A}\ }\textbf {\bibinfo {volume} {56}},\ \bibinfo {pages} {1638} (\bibinfo {year} {1997})}\BibitemShut {NoStop}%
\bibitem [{\citenamefont {Pati}\ and\ \citenamefont {Lawande}(1998)}]{Pati_1998}%
  \BibitemOpen
  \bibfield  {author} {\bibinfo {author} {\bibfnamefont {A.~K.}\ \bibnamefont {Pati}}\ and\ \bibinfo {author} {\bibfnamefont {S.~V.}\ \bibnamefont {Lawande}},\ }\href {https://doi.org/10.1103/PhysRevA.58.831} {\bibfield  {journal} {\bibinfo  {journal} {Physical Review A}\ }\textbf {\bibinfo {volume} {58}},\ \bibinfo {pages} {831} (\bibinfo {year} {1998})}\BibitemShut {NoStop}%
\bibitem [{\citenamefont {Pati}(1999)}]{Pati1999}%
  \BibitemOpen
  \bibfield  {author} {\bibinfo {author} {\bibfnamefont {A.~K.}\ \bibnamefont {Pati}},\ }\href {http://www.physics.sk/aps/pub.php?y=1999&pub=aps-99-04} {\bibfield  {journal} {\bibinfo  {journal} {Acta Physica Slovaca}\ }\textbf {\bibinfo {volume} {49}},\ \bibinfo {pages} {567} (\bibinfo {year} {1999})}\BibitemShut {NoStop}%
\bibitem [{\citenamefont {Fischer}\ \emph {et~al.}(2001)\citenamefont {Fischer}, \citenamefont {Guti\'errez-Medina},\ and\ \citenamefont {Raizen}}]{Fischer2001}%
  \BibitemOpen
  \bibfield  {author} {\bibinfo {author} {\bibfnamefont {M.~C.}\ \bibnamefont {Fischer}}, \bibinfo {author} {\bibfnamefont {B.}~\bibnamefont {Guti\'errez-Medina}},\ and\ \bibinfo {author} {\bibfnamefont {M.~G.}\ \bibnamefont {Raizen}},\ }\href {https://doi.org/10.1103/PhysRevLett.87.040402} {\bibfield  {journal} {\bibinfo  {journal} {Physical Review Letters}\ }\textbf {\bibinfo {volume} {87}},\ \bibinfo {pages} {040402} (\bibinfo {year} {2001})}\BibitemShut {NoStop}%
\bibitem [{\citenamefont {Tidstr\"om}\ and\ \citenamefont {Sj\"oqvist}(2003)}]{Jonas2003}%
  \BibitemOpen
  \bibfield  {author} {\bibinfo {author} {\bibfnamefont {J.}~\bibnamefont {Tidstr\"om}}\ and\ \bibinfo {author} {\bibfnamefont {E.}~\bibnamefont {Sj\"oqvist}},\ }\href {https://doi.org/10.1103/PhysRevA.67.032110} {\bibfield  {journal} {\bibinfo  {journal} {Physical Review A}\ }\textbf {\bibinfo {volume} {67}},\ \bibinfo {pages} {032110} (\bibinfo {year} {2003})}\BibitemShut {NoStop}%
\bibitem [{\citenamefont {de~Faria}\ \emph {et~al.}(2003)\citenamefont {de~Faria}, \citenamefont {de~Toledo~Piza},\ and\ \citenamefont {Nemes}}]{Faria_2003}%
  \BibitemOpen
  \bibfield  {author} {\bibinfo {author} {\bibfnamefont {J.~G.~P.}\ \bibnamefont {de~Faria}}, \bibinfo {author} {\bibfnamefont {A.~F.~R.}\ \bibnamefont {de~Toledo~Piza}},\ and\ \bibinfo {author} {\bibfnamefont {M.~C.}\ \bibnamefont {Nemes}},\ }\href {https://doi.org/10.1209/epl/i2003-00440-4} {\bibfield  {journal} {\bibinfo  {journal} {Europhysics Letters}\ }\textbf {\bibinfo {volume} {62}},\ \bibinfo {pages} {782} (\bibinfo {year} {2003})}\BibitemShut {NoStop}%
\bibitem [{\citenamefont {Schmidt}(2004)}]{Schmidt2004}%
  \BibitemOpen
  \bibfield  {author} {\bibinfo {author} {\bibfnamefont {A.~U.}\ \bibnamefont {Schmidt}},\ }\href {https://doi.org/10.48550/arXiv.math-ph/0307044} {\bibfield  {journal} {\bibinfo  {journal} {arXiv:0307044}\ } (\bibinfo {year} {2004})}\BibitemShut {NoStop}%
\bibitem [{\citenamefont {Itano}(2009)}]{Itano_2009}%
  \BibitemOpen
  \bibfield  {author} {\bibinfo {author} {\bibfnamefont {W.~M.}\ \bibnamefont {Itano}},\ }\href {https://doi.org/10.1088/1742-6596/196/1/012018} {\bibfield  {journal} {\bibinfo  {journal} {Journal of Physics: Conference Series}\ }\textbf {\bibinfo {volume} {196}},\ \bibinfo {pages} {012018} (\bibinfo {year} {2009})}\BibitemShut {NoStop}%
\bibitem [{\citenamefont {Jacobs}(2014)}]{Jacobs2014}%
  \BibitemOpen
  \bibfield  {author} {\bibinfo {author} {\bibfnamefont {K.}~\bibnamefont {Jacobs}},\ }\href {https://doi.org/https://doi.org/10.1017/CBO9781139179027} {\emph {\bibinfo {title} {Quantum Measurement Theory and its Applications}}}\ (\bibinfo  {publisher} {Cambridge University Press},\ \bibinfo {year} {2014})\BibitemShut {NoStop}%
\bibitem [{\citenamefont {Kulaga}(1995)}]{Kulaga1995}%
  \BibitemOpen
  \bibfield  {author} {\bibinfo {author} {\bibfnamefont {A.}~\bibnamefont {Kulaga}},\ }\href {https://doi.org/https://doi.org/10.1016/0375-9601(95)00278-B} {\bibfield  {journal} {\bibinfo  {journal} {Physics Letters A}\ }\textbf {\bibinfo {volume} {202}},\ \bibinfo {pages} {7} (\bibinfo {year} {1995})}\BibitemShut {NoStop}%
\bibitem [{\citenamefont {García-Pintos}\ and\ \citenamefont {del Campo}(2019)}]{Pintos_2019}%
  \BibitemOpen
  \bibfield  {author} {\bibinfo {author} {\bibfnamefont {L.~P.}\ \bibnamefont {García-Pintos}}\ and\ \bibinfo {author} {\bibfnamefont {A.}~\bibnamefont {del Campo}},\ }\href {https://doi.org/10.1088/1367-2630/ab099e} {\bibfield  {journal} {\bibinfo  {journal} {New Journal of Physics}\ }\textbf {\bibinfo {volume} {21}},\ \bibinfo {pages} {033012} (\bibinfo {year} {2019})}\BibitemShut {NoStop}%
\bibitem [{\citenamefont {Pati}(1995)}]{Pati1995}%
  \BibitemOpen
  \bibfield  {author} {\bibinfo {author} {\bibfnamefont {A.~K.}\ \bibnamefont {Pati}},\ }\href {https://doi.org/https://doi.org/10.1016/0375-9601(95)00299-I} {\bibfield  {journal} {\bibinfo  {journal} {Physics Letters A}\ }\textbf {\bibinfo {volume} {202}},\ \bibinfo {pages} {40} (\bibinfo {year} {1995})}\BibitemShut {NoStop}%
\bibitem [{\citenamefont {Brody}\ and\ \citenamefont {Graefe}(2012)}]{Brody2012}%
  \BibitemOpen
  \bibfield  {author} {\bibinfo {author} {\bibfnamefont {D.~C.}\ \bibnamefont {Brody}}\ and\ \bibinfo {author} {\bibfnamefont {E.-M.}\ \bibnamefont {Graefe}},\ }\href {https://doi.org/10.1103/PhysRevLett.109.230405} {\bibfield  {journal} {\bibinfo  {journal} {Physical Review Letters}\ }\textbf {\bibinfo {volume} {109}},\ \bibinfo {pages} {230405} (\bibinfo {year} {2012})}\BibitemShut {NoStop}%
\bibitem [{\citenamefont {Cimmarusti}\ \emph {et~al.}(2015)\citenamefont {Cimmarusti}, \citenamefont {Yan}, \citenamefont {Patterson}, \citenamefont {Corcos}, \citenamefont {Orozco},\ and\ \citenamefont {Deffner}}]{Cimmarusti2015}%
  \BibitemOpen
  \bibfield  {author} {\bibinfo {author} {\bibfnamefont {A.~D.}\ \bibnamefont {Cimmarusti}}, \bibinfo {author} {\bibfnamefont {Z.}~\bibnamefont {Yan}}, \bibinfo {author} {\bibfnamefont {B.~D.}\ \bibnamefont {Patterson}}, \bibinfo {author} {\bibfnamefont {L.~P.}\ \bibnamefont {Corcos}}, \bibinfo {author} {\bibfnamefont {L.~A.}\ \bibnamefont {Orozco}},\ and\ \bibinfo {author} {\bibfnamefont {S.}~\bibnamefont {Deffner}},\ }\href {https://doi.org/10.1103/PhysRevLett.114.233602} {\bibfield  {journal} {\bibinfo  {journal} {Phys. Rev. Lett.}\ }\textbf {\bibinfo {volume} {114}},\ \bibinfo {pages} {233602} (\bibinfo {year} {2015})}\BibitemShut {NoStop}%
\bibitem [{\citenamefont {Deffner}(2022)}]{Deffner_2022}%
  \BibitemOpen
  \bibfield  {author} {\bibinfo {author} {\bibfnamefont {S.}~\bibnamefont {Deffner}},\ }\href {https://doi.org/10.1209/0295-5075/ac9fed} {\bibfield  {journal} {\bibinfo  {journal} {Europhysics Letters}\ }\textbf {\bibinfo {volume} {140}},\ \bibinfo {pages} {48001} (\bibinfo {year} {2022})}\BibitemShut {NoStop}%
\end{thebibliography}%

\section{Appendix}\label{detailedproof}
The variances of the free Hamiltonian $H_0$ and the part due to measurement, namely, $H_1$ in the state $\ket{\Psi(t)}$ are
\begin{widetext}
\begin{eqnarray}\label{varH0andH1}
\Delta H_0^2 &=&\frac{\sum_i\exp[-2f\int g(\frac{a_i-\tilde{a}(t)}{\Delta \tilde{a}})dt]E_{0i}^2 \abs{\Phi_i(0)}^2}{\sum_i\exp[-2f\int g(\frac{a_i-\tilde{a}(t)}{\Delta \tilde{a}})dt]\abs{\Phi_i(0)}^2}-\left(\frac{\sum_i\exp[-2f\int g(\frac{a_i-\tilde{a}(t)}{\Delta \tilde{a}})dt]E_{0i} \abs{\Phi_i(0)}^2}{\sum_i\exp[-2f\int g(\frac{a_i-\tilde{a}(t)}{\Delta \tilde{a}})dt]\abs{\Phi_i(0)}^2}\right)^2,\nonumber\\
\Delta H_1^2 &=&\frac{\sum_i\exp[-2f\int g(\frac{a_i-\tilde{a}(t)}{\Delta \tilde{a}})dt](\hbar f g(\frac{a_i-\tilde{a}(t)}{\Delta \tilde{a}}))^2 \abs{\Phi_i(0)}^2}{\sum_i\exp[-2f\int g(\frac{a_i-\tilde{a}(t)}{\Delta \tilde{a}})dt]\abs{\Phi_i(0)}^2}-\left(\frac{\sum_i\exp[-2f\int g(\frac{a_i-\tilde{a}(t)}{\Delta \tilde{a}})dt]\hbar f g(\frac{a_i-\tilde{a}(t)}{\Delta \tilde{a}}) \abs{\Phi_i(0)}^2}{\sum_i\exp[-2f\int g(\frac{a_i-\tilde{a}(t)}{\Delta \tilde{a}})dt]\abs{\Phi_i(0)}^2}\right)^2.\nonumber\\
\end{eqnarray}
\end{widetext}
Let us now consider the approximation $\exp[-2f\int g(\frac{a_i-\tilde{a}(t)}{\Delta \tilde{a}})dt]\approx 1-2f\int g(\frac{a_i-\tilde{a}(t)}{\Delta \tilde{a}})dt$ then the expectation value of $H_0$ in the state $\ket{\Psi(t)}$ is
\begin{eqnarray}
\langle H_0 \rangle_t =\frac{\sum_i(1-2f\int g(\frac{a_i-\tilde{a}(t)}{\Delta \tilde{a}})dt)E_{0i} \abs{\Phi_i(0)}^2}{1-2f\sum_i\int g(\frac{a_i-\tilde{a}(t)}{\Delta \tilde{a}})dt \abs{\Phi_i(0)}^2},
\end{eqnarray}
where in the denominator we have used the fact that $\sum_i \abs{\Phi_i(0)}^2=1$. For simplicity, we assume that the measurement results are constant in time $\tilde{a}(t)=\tilde{a}$ for all $t$ then $\langle H_0 \rangle_t$ becomes
\begin{eqnarray}
   \langle H_0 \rangle_t =\frac{\langle H_0 \rangle_0-2tf\langle g H_0 \rangle_0}{1-2tf\langle g \rangle_0}, 
\end{eqnarray}
where $\langle H_0 \rangle_0=\sum_i E_{0i}\abs{\Phi_i(0)}^2$ and $\langle g \rangle_0=\sum_i g(\frac{a_i-\tilde{a}}{\Delta \tilde{a}})\abs{\Phi_i(0)}^2$. We can now Taylor expand the denominator of the above equation to get
\begin{eqnarray}
   \langle H_0 \rangle_t =\langle H_0 \rangle_0+2tf\langle H_0 \rangle_0\langle g \rangle_0-2tf\langle g H_0 \rangle_0, 
\end{eqnarray}
where we have used the same approximation as above, i.e., we neglect terms of the order $\mathcal{O}(f^2g^2)$.
Similarly, we obtain
\begin{eqnarray}
   \langle H_0^2 \rangle_t =\langle H_0^2 \rangle_0+2tf\langle H_0^2 \rangle_0\langle g \rangle_0-2tf\langle g H_0^2 \rangle_0. 
\end{eqnarray}
Hence, the variance of $H_0$ under this approximation is
\begin{eqnarray}
   \Delta H_0^2(t)&=& \Delta H_0^2(0)+2tf[ \langle H_0^2 \rangle_0 \langle g \rangle_0- \langle g H_0^2 \rangle_0\nonumber\\
   &&-2{\langle H_0 \rangle_0}^2 \langle g \rangle_0+2\langle g H_0 \rangle_0 \langle H_0 \rangle_0].
\end{eqnarray}
It is obvious that the variance of $H_1$ is at least of the order $\mathcal{O}(f^2g^2)$ (see Eq.\eqref{varH0andH1}), so it vanishes under this approximation. Therefore, using Eq.\eqref{speed} and the fact that $[H_1,H_0]=0$, we obtain
\begin{eqnarray}
    V=\frac{2}{\hbar}\sqrt{\Delta H_0^2(0)+2tfX},
\end{eqnarray}
where $X:=2\langle H_0 \rangle_0 \text{Cov}(g,H_0)-\text{Cov}(g,H_0^2)$, and $\text{Cov}(A,B)=\langle AB \rangle-\langle A \rangle\langle B \rangle$ is the covariance of $A$ and $B$.

\end{document}